\documentclass{elsart}
\usepackage{epsfig}
\usepackage{amsmath}
\usepackage{amssymb}
\usepackage{graphicx}
\usepackage{subfigure}
\usepackage{color}

\newcommand \be{\begin{equation}}
\newcommand \ba{\begin{eqnarray}}
\newcommand \ee{\end{equation}}
\newcommand \ea{\end{eqnarray}}

\begin{document}
\runauthor{Zhou and Sornette}
\begin{frontmatter}
\title{Causal Slaving of the U.S. Treasury Bond Yield Antibubble
by the Stock Market Antibubble of August 2000}
\author[iggp]{\small{Wei-Xing Zhou}},
\author[iggp,ess,nice]{\small{Didier Sornette}\thanksref{EM}}
\address[iggp]{Institute of Geophysics and Planetary Physics, University of
California, Los Angeles, CA 90095}
\address[ess]{Department of Earth and Space Sciences, University of
California, Los Angeles, CA 90095}
\address[nice]{Laboratoire de Physique de la Mati\`ere Condens\'ee,
CNRS UMR 6622 and Universit\'e de Nice-Sophia Antipolis, 06108 Nice Cedex 2, France}
\thanks[EM]{Corresponding author. Department of Earth and Space
Sciences and Institute of Geophysics and Planetary Physics,
University of California, Los Angeles, CA 90095-1567, USA. Tel:
+1-310-825-2863; Fax: +1-310-206-3051. {\it E-mail address:}\/
sornette@moho.ess.ucla.edu (D. Sornette)\\
http://www.ess.ucla.edu/faculty/sornette/}

\begin{abstract}
Using the descriptive method of log-periodic power laws (LPPL)
based on a theory of behavioral herding, we use a battery of
parametric and non-parametric tests to demonstrate the
existence of an antibubble in the yields with maturities larger than 1 year
since October 2000. The concept of ``antibubble'' describes the
existence of a specific LPPL pattern that is thought to reflect
collective herding effects. From the dependence of the parameters of the LPPL formula
as a function of yield maturities and using lagged cross-correlation
calculations between the S\&P 500 and bond yields, we find strong
evidence for the following causality:
Stock Market $\to$ Fed Reserve (Federal
funds rate) $\to$ short-term yields $\to$ long-term yields (as well
as a direct and instantaneous influence of the stock market on the long-term yields).
Our interpretation is that the FRB is ``causally slaved'' to the stock market
(at least for the studied period),
because the later is (taken as) a proxy for the present and future health of
the economy.
\end{abstract}


\end{frontmatter}

\typeout{SET RUN AUTHOR to \@runauthor}

\section{Introduction}
\label{s1:introduction}

Since August 2000 until the end of 2002,
the USA as well as most other western markets have
depreciated almost in synchrony according to complex patterns of drops
and local rebounds. We have proposed to describe this phenomenon using
the concept of a log-periodic power law (LPPL) ``antibubble,''
characterizing behavioral herding between investors leading to a
competition between positive and negative feedbacks in the pricing
process \cite{SZ02QF,ZS03RG,J03QF,SZ03QF,ZS03PAGlobal}.
The concept of an ``antibubble'' was inspired by that of an ``antiparticle'' in
physics. Just as an antiparticle is identical to its sister particle
except that it carries exactly opposite charges and destroys its sister
particle upon encounters, an antibubble is both the same and the
opposite of a bubble; it's the same because similar herding patterns
occur, but with a bearish vs. bullish slant. Our work on antibubbles
is thus the counterpart of a large research effort that we and others
have developed to characterize speculative bubbles (see
\cite{SJ01sign,S03,S03PR,JS04endo} and references therein).

However, in addition to imitative and herding behavior, one
quantifiable factor that many analysts and traders think plays an
important role in the development of both bubbles and antibubbles
is the influence of the Federal Reserve Board's (FRB) liquidity
interventions into the financial markets.  On a daily basis, the
Federal Reserve intervenes to adjust short-term interest rates.
Through open market operations, the FRB buys and sells U.S.
Government securities (variously-termed Treasury instruments
having term periods that range from overnight to a couple of
weeks) in the secondary market in order to adjust the level of
reserves in the banking system. By adjusting the level of reserves
in the banking system through their buys or sells which modulate
the supply-demand equation, the FRB can offset or support seasonal
or cyclical shifts of funds and thereby affect short-term interest
rates and the growth of the money supply, thus effecting the cost
of borrowing in the larger economy.ÊÊ

The impact of the FRB does not stop with the short-term interest
rates but spreads to bonds (mechanically) and to the stock markets
through several channels. One particularly simple channel is the
influence of the risk-free interest rates on the present value of
discounted future earnings and dividends. The interest rates, as
the prices or costs of money at different time horizons, are
affected by production opportunities, time preference for
consumption, risk and expected inflation. The FRB uses a model of
the U.S. economy to shape its decisions that takes into account
the monetary transmission mechanism, that is, how its monetary
policy actions influence financial markets and aggregate output
and inflation \cite{FRBmodel}. By its interventions, the FRB puts
or removes short-term liquidity into the hands of bankers who use
it to invest in or desinvest from various financial instruments
including equities.  This would obviously have an influence on
near-term market move directions. One can argue that the FRB
intervenes in the stock markets by this indirect mean. An example
of extraordinary action by the FRB was in the months prior to the
turn of the Millennium.  At the time, the Fed reported their
intent to ameliorate possible liquidity problems on January 1,
2000 by the injection of several hundred billion dollars.  Many
believe that this liquidity was a substantial factor in early
doubling the value of the Nasdaq stock market into the first
quarter of 2000. Another extreme example is, in the week following
September 11, 2001, when the Fed reported Open Market Operations
activity of \$76 billion, one of the single largest weekly
liquidity doses in recent history. In August 2003, in response to
the power outages of the Northeast and Canada, the FRB's market
intervention peaked to \$48 billion in a single day in total
outstanding repurchase agreements, an amount approximately equal
to that needed to rebuild the flawed electrical power grid in
America. Based on comparison between the net value of outstanding
Fed repurchase agreements (repo) and financial indices such as the
S\&P 500 index, some analysts have suggested a correlation between
the dips of the market and the ramping up of FRB repo activities,
implying a causal relationship in which the FRB influences the
stock market (see for instance
http://www.piraz.com/monetary/temp99b.gif). An interesting
suggestion is that the FRB would appear to inject liquidity to the
market not in a steady--state fashion but only at those times when
its action may have a strong impact. The rational for this is
based on the recognition that many Americans engaged themselves
economically through the financial markets in the late 1990s and
the capitalization of the US stock market is now significantly
larger than the US GDP. There is temptation to view the stock
markets as leading the economy. Knowledgeable investors and
traders then understand that the FRB acts accordingly to use
liquidity to push the markets and, thus, push the economy.

In contrast to this arrow of causality in which it is the FRB
which influences the stock market, others suggest the opposite
direction of causality, namely, that the monetary policy reacts to
the stock markets. However, the magnitude of the FRB's reaction to
the stock market is very hard to estimate in part because of the
simultaneous response of equity prices to interest rates
\cite{RS03QJE}. The near-term path of the FRB's policy might respond
to equity price movements due to the direct wealth effect (a
higher stock market spill over to increase consumption) and
because the stock market prices might contain information on the
future economic activity (the current prices are thought to
reflect expectations of future earnings and dividends). Rigobon
and Sack have found evidence for a significant monetary policy
response to the stock market, using an identification technique
based on the heteroskedasticity of the stock market returns
applied to daily S\&P 500 index data running from March 1985 to
December 1999 \cite{RS03QJE}. Their method consists in
conditioning the covariance matrix between stock market returns
and the three-month treasury bill interest rate on different
volatility levels in order to identify the impact of volatility
regimes. Bohl, Siklos and Werner \cite{BSW03SSRN} argued that the
Bundesbank reacted to the stock market systematically on one month
scale (or lower frequency), although the effect is not
significant.

This question is embedded within a broader question: what is the
information content of interest rate time series and of the shape
of the yield curve \footnote{The shapes and deformations of the
yield curve, giving interest rates as a function of maturities,
can be explained by several theories such as investor's
expectation about future inflation rates, liquidity preference and
market segmentation based on the competition between the supply
and demand of long-term and short-term treasury securities.
Therefore, the interest rate spread can be utilized as an
indicator of the future real economic activities and of inflation.
The average parabolic shape of the yield curve as a function of
maturity can be explained by a value-at-risk argument on expected
future risks \cite{BSCEP99AMF,MB00IJTAFa,MB00IJTAFb}.}?
Empirically, it has been shown that interest rate
spreads\footnote{the difference between a long-term interest rate
and a short-term interest rate} have predictive power for the real
economic activity and for inflation. For many countries, the
interest spreads seem to outperform other leading financial
indicators
\cite{EM97EER,EM98RES,ILS00IJF,BOS01SJPE,A02JIMF,ERS02RES,HK02JMCB,H03CJE}.
This is over controversial: financial variables are sometimes
argued not to predict real activities \cite{TG98EI,W02EI} while
leading economic indicators do have predictive power
\cite{Q01IJF,CP02JAE}. There is also evidence that the interest
rates and their spreads variables could potentially be used as
leading indicators for real estate markets \cite{BT01AE}. Yield
spreads have been used successfully to forecast the US recession
in 2001 \cite{CP02EL,S03JRS,M01XXX}, which occurred about a year
after the ``new economy'' bubble burst. With respect to the stock
market, Roehner has found strong negative correlations between
stock market crash-recovery and interest rate spread
\cite{R00IJMPC}, while Resnick and Shoesmith showed that the
interest rates spread between the yields of US composite 10-year+
US T-bond and three-month T-bill carries informative content of a
forthcoming bear market one month in advance \cite{RS02FAJ}. In
contrast, the interest rates (or bond yields) have been argued to
have no significant forecasting ability \cite{F92}. This suggests
that the arrow of causality goes from the interest rates to the
stock market, consistent with the above idea that the FRB may
influence the stock market through its monetary policy.

Here, we provide a novel approach to this topic by analyzing the
FRB's reaction to the stock market after the 2000 crash. Our
methodology is more efficient than previous works because we use a
first-order deterministic indicator rather than second-order
covariance measures. Our study also capitalizes on the rather
unique setting since 2000 provided by the stock market on the one
hand and by the FRB's monetary policy on the other hand and on our
ability to model quantitatively these behaviors. First, we
identify specific log-periodic power law (LPPL) signatures in the
treasury security yields. Specifically, Sec.~\ref{s1:YiledAB}
presents our LPPL analysis of the US treasury securities yields,
which shows that an antibubble also started in October 2000 on the
yields with maturities of 2 years or larger. Section
\ref{s2:harmonic} strengthens the evidence for log-periodicity by
demonstrating the existence of a strong third harmonic in the
log-periodicity. Section \ref{s2:Hierarchy} shows the existence of
antibubbles within antibubbles in the time dependence of the
long-maturity yields since 1979. Section \ref{s1:LPinFFR}
demonstrates the existence of a log-periodic pattern in the timing
of the moves of the Federal funds rate. In other words, the times
at which the FRB lowered its leading indicator are organized
according to an approximate geometric series with accumulation
time (going backward) around October 2000. Section \ref{s1:Causality} shows that
the pattern of timing of the FRB action closely matches that of
significant drops in the stock market with an average time lag of
about 1-2 months. In addition, the critical time $t_c$ of the antibubble
LPPL patterns of the interest rate time series are found all
approximately three months after that the critical time of the
stock market antibubble. This strongly suggests that the FRB's
actions reacted to the stock market rather than the reverse. Section \ref{s1:concude}
concludes.

\section{Evidence of an antibubble structure in the time
evolution of U.S. Treasury bond yields}
\label{s1:YiledAB}

\subsection{The LPPL antibubble framework and its calibration} \label{s2:Theory}

We use the theory characterizing behavioral herding between
investors in terms of a competition between positive and negative
feedbacks in the pricing process, which has been documented in
\cite{SJ01sign,S03,S03PR}. Specifically, we refer to the
parametrization \cite{GS02PRE,ZS03RG} adapted to the description
of antibubbles under the form: \be y\left( t\right) = A +
\left(t-t_c \right) ^{m} \left\{B+ \sum_{n=1}^N C_n \cos\left[n
\omega \ln \left(t-t_c \right) - \phi_n \right] \right\} +
\epsilon(t)~, \label{Eq:LPPL} \ee where $t_c$ is the theoretical
inception time of the antibubble and $\epsilon(t)$ is a white
noise residue. The first-order ($N=1$) and second-order ($N=2$)
formulae are the versions that have been most used previously in
modelling and predicting financial bubbles and antibubbles. We
refer to Ref.~\cite{S03,S03PR} and references therein for a full
exposition of the approach.

\begin{figure}
\begin{center}
\includegraphics[width=7cm]{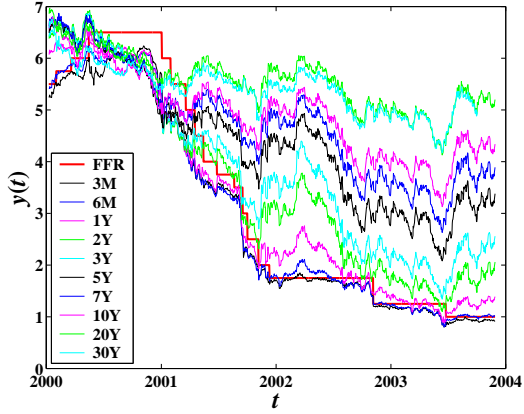}
\end{center}
\caption{(Color online) Time evolution of ten yields with the
maturities 3M (three months), 6M, 1Y (one year), 2Y, 3Y, 5Y, 7Y,
10Y, 20Y, 30Y and of the Federal funds rate since January 2000.}
\label{Fig:FfrYields3}
\end{figure}

Figure \ref{Fig:FfrYields3} shows the evolution with times of ten
yields with the maturities 3M, 6M, 1Y, 2Y, 3Y, 5Y, 7Y, 10Y, 20Y,
30Y and of the Federal funds rate\footnote{The federal funds rate
is the interest rate at which depository institutions lend
balances at the Federal Reserve to other depository institutions
overnight.} since January 2000. As can be seen from the almost
simultaneous crossing of all yields in October 2000 which is close
to the inception of the 2000 yield antibubble documented below,
the yield curve exhibited a rather anomalous inverted shape during
the first half of 2000 and resumed its normal upward concave shape
after, with a spread between the rates of different maturities
broadening with time.

\begin{figure}[h]
\begin{center}
\includegraphics[width=7cm]{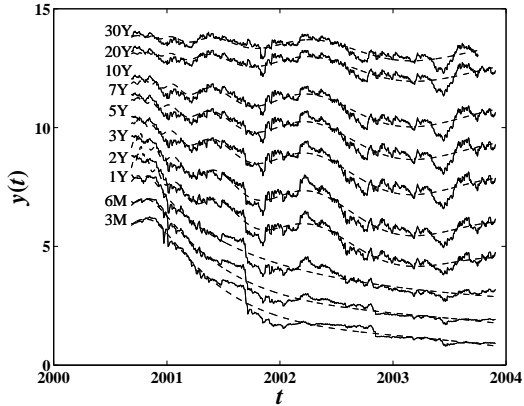}
\end{center}
\caption{Time series (wiggly lines) of the yields with ten
different maturities (3M, 6M, 1Y, 2Y, 3Y, 5Y, 7Y, 10Y, 20Y, 30Y)
and their fits (dashed lines) using the first order ($N=1$) LPPL
formula (\ref{Eq:LPPL}). The parameters of the fits are listed in
Table \ref{Tb:LPPL1}. The yields with maturities greater than
three months have been shifted vertically upwards by 0.9\%
incrementally.} \label{Fig:YieldFit}
\end{figure}

Figure \ref{Fig:YieldFit} and Table \ref{Tb:LPPL1} show the
results of the fits of the ten yields with different maturities
(3M, 6M, 1Y, 2Y, 3Y, 5Y, 7Y, 10Y, 20Y, 30Y) shown in
Fig.~\ref{Fig:FfrYields3} with the first-order ($N=1$) LPPL
formula (\ref{Eq:LPPL}). The yields with maturities no less than
2Y are similar to each other with approximately synchronous
log-periodic oscillations. Yields with short maturities (3M, 6M,
and 1Y) exhibit different shapes. The parameters of the fits of
(\ref{Eq:LPPL}) with $N=1$ to the yields are listed in Table
\ref{Tb:LPPL1}.

Several interesting features can be
extracted from this analysis. First of all, we observe a universal
log-periodic oscillatory structure for yields with maturities no
less than 2Y, with approximately the same angular log-frequencies
$\omega \approx 7.4$, as given in Table \ref{Tb:LPPL1}.
In contrast, for the yields with maturities 3M, 6M and 1Y,
the small values of
$\omega$ and the values of the exponents $m$ close to $1$
(absence of curvature) or to $0$ (curvature localized close to $t_c$)
indicate the absence of LPPL.

\begin{table}
\begin{center}
\caption{\label{Tb:LPPL1} Parameters of the fits with the
first-order LPPL formula (\ref{Eq:LPPL}) with $N=1$ off the ten
U.S. Treasury bond yields shown in Fig.~\ref{Fig:FfrYields3}.}
\medskip
\begin{tabular}{ccccccccccccc}
\hline\hline
Maturity&$t_c$&$m$&$\omega$&$\phi_1$&$A$&$B$&$C_1$&$\chi$\\\hline
3M&2000/10/22&1.00&1.40&4.13&6.186&-0.00823& 0.00408& 0.206\\
6M&2000/10/19&0.98&1.29&3.42&6.132&-0.00884& 0.00454& 0.199\\
1Y&2000/10/15&0.15&0.85&2.80&8.411&-1.97804& 0.55760& 0.230\\
2Y&2000/10/08&0.42&7.94&6.23&6.855&-0.28056& 0.02946& 0.224\\
3Y&2000/10/07&0.49&7.73&4.84&6.366&-0.14033& 0.02216& 0.249\\
5Y&2000/10/01&0.64&7.60&4.01&5.883&-0.03523& 0.00861& 0.262\\
7Y&2000/09/29&0.67&7.43&2.95&5.889&-0.02138& 0.00639& 0.247\\
10Y&2000/10/21&0.64&7.06&0.26&5.730&-0.01959& 0.00805& 0.236\\
20Y&2000/10/19&0.75&6.87&5.40&5.965&-0.00452& 0.00295& 0.205\\
30Y&2000/09/28&0.80&7.52&3.61&5.775&-0.00289& 0.00179& 0.187\\
\hline\hline
\end{tabular}
\end{center}
\end{table}

Secondly, the estimated inception dates $t_c$ for the yield
antibubbles are consistent across all maturities to within three
weeks, approximately in the first half of October 2000. This
places the inception of the yield antibubble about two months
after that of the worldwide antibubble (first half of August 2000)
\cite{SZ02QF,ZS03RG,SZ03QF,ZS03PAGlobal}. In spite of the absence
of a well-developed log-periodic structure for the 3-month yield,
the power law still provides a reasonable estimation of $t_c$.

Thirdly, the progressive return of the yield curve to its normal upward concave
shape after October 2000 is reflected in
the decrease of the absolute value of the coefficient $B$ with maturity:
a larger $|B|$ for the smaller maturities imply that the corresponding
yields are decaying faster as a function of time. This leads to a growing spread and quantifies
the evolution of the yield curve from a basically flat shape in October 2000
to a normal upward concave shape later. We find that the coefficient $|B|$
for a given yield maturity $\theta$ is approximately
inversely proportional to the inverse of the square of the maturity:
\be
|B| \sim {1 \over \theta^2}~. \label{flallw}
\ee
This dependence is different from that derived from the square-root
law proposed in \cite{BSCEP99AMF,MB00IJTAFa,MB00IJTAFb} to describe the normal regime,
based on a Value-at-Risk like pricing
of the forward rate curve. This is an additional evidence suggesting that the antibubble
corresponds to an anomalous (or at least different) regime.

\subsection{Statistical significance of the log-periodic structure}
\label{s2:SigLev}

\begin{figure}[h]
\begin{center}
\includegraphics[width=7cm]{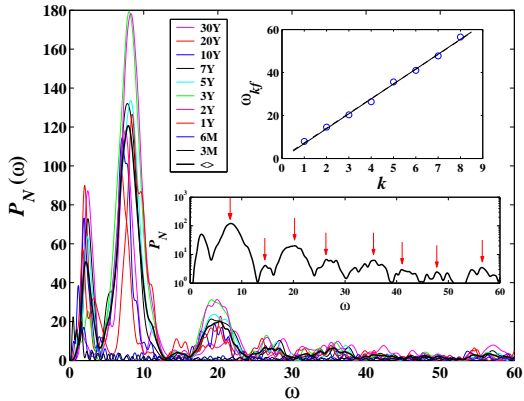}
\end{center}
\caption{(Color online) Lomb periodograms of the detrended
residuals of the ten treasury security yields and their Lomb
average shown as the thickest line (referd to in the legend as
``$\langle \rangle$''). Except for the 3M and 6M yields, all the
other eight yield residues exhibit a significant Lomb peak at
approximately the same fundamental angular log-frequency on on
several of its harmonics. The lower inset shows the averaged Lomb
periodogram in semi-log plot. The vertical arrows point to the
peaks that identify a series of characteristic angular
log-frequencies $\omega_{kf}$'s used to construct the upper inset.
The upper inset presents the relation between $\omega_{kf}$ and
$k$ for $k=1,2,\cdots,8$. The solid line shows an unconstrained
linear regression, while the dashed line is the linear regression
with the constraint of passing through the origin.}
\label{Fig:Lomb}
\end{figure}

To assess the significance level of the observed log-periodic
pattern, we present several statistical tests based on spectral
analysis. The standard test consists in performing a Lomb
periodogram analysis of the de-trended time series
\begin{equation}
S(\ln(t)) = \frac{y(t)-A}{(t-t_c)^{m}}~, \label{Eq:Res}
\end{equation}
in the manner introduced in Ref.~\cite{JLS00} in a similar context.
The Lomb periodogram analysis is
analogous to a Fourier transform but for unevenly spaced data
\cite{Press}. If $y(t)$ possesses log-periodic power law
structure, $S$ has regular cosine undulations in the variable $\ln(t)$.
Figure \ref{Fig:Lomb} shows the ten Lomb periodograms of the
detrended signal of the ten treasury security yields. Except for the
3M and 6M yields, all the other eight yields exhibit a significant
Lomb peak at the same fundamental angular log-frequency $\omega$ and also
present peaks at harmonics $n\omega$ (where $n$ is an integer).
It is interesting to observe that the parametric fit of the
1Y yield with formula (\ref{Eq:LPPL}) fails to qualify a LPPL structure
while the present non-parametric method unearths a clear log-periodic structure. The
two highest peaks are found for the maturities 2Y and 3Y. The thickest line shows
the average of the Lomb periodograms over the eight periodograms of the yields
with maturities larger than 6M. This averaging procedure has been
introduced in \cite{JS98IJMPC} as an efficient way of enhancing a periodic
or log-periodic signal when one has the luxury of an ensemble of realizations.
Our present average amounts to postulate that each of the eight maturities
larger than 6M has the same log-periodicity but with different residual noises.

The average normalized Lomb periodogram has its highest peak equal
to $P_N(\omega_f) = 121$ at the fundamental angular log-frequency
$\omega_f=7.92$. The false-alarm probability under the null
hypothesis of i.i.d. Gaussian fit residuals is zero \cite{Press}.
But assuming an i.i.d. structure is too restrictive (too
optimistic) as the fit residuals have a dependence. We assume that
this dependence can be approximated by the model of fractional
Brownian noise with a Hurst exponent $H>1/2$. We can use our
previous construction of a table of false alarm probabilities for
various values of $H$ \cite{ZS02IJMPC}. For $H=0.6$, the false
alarm probability corresponding to the observed peak
$P_N(\omega_f) = 121$ is $<0.01\%$, for $H=0.7$ it is $0.2\%$, for
$H=0.8$ it is $3\%$ and for $H=0.9$ it is $20\%$. This means that
the statistical significance of the log-periodicity is very high.

Bothmer has addressed the problem of the influence of noise
dependence in the determination of the statistical significance of
log-periodic oscillations \cite{B03QF}. He considers specifically
the dependence introduced in data which have a cumulative nature
(like a price which is the logarithm of the sum of returns). For
this, Bothmer introduced the so-called cumulative Lomb periodogram
and found it to be exponentially distributed independently of the
frequency \cite{B03QF} for the null hypothesis of no oscillations.
Actually, using a data set which involves a sum of noise
contributions around a power law leads to a spurious peak on the
Lomb periodogram corresponding to a most probable noise
\cite{HJLSS00}. This mechanism explains the peak at
$\omega_{\rm{mp}}=2.42$ observed in the averaged Lomb periodogram,
which corresponds to about $1-1.5$ oscillations over the whole
time span.

A strong additional evidence for log-periodicity
lies on the presence of several
higher-order harmonics of the fundamental $\omega_f$, as
illustrated in the lower inset of Fig.~\ref{Fig:Lomb}.
Here, we follow our previous aproach applied to hydrodynamic turbulence
\cite{ZS02PD,ZSP03IJMPC}.
The vertical arrows point to the peaks that identify a series of characteristic
angular log-frequencies $\omega_{kf}$'s as a function of their order $k$ from
left to right. The upper inset of Fig.~\ref{Fig:Lomb}
shows the linear relationship between the $\omega_{kf}$'s as a function
of their order $k$ for $k=1,2,\cdots,8$. A linear regression is shown
as the dashed line and gives
$\omega_{kf} = 6.87 k + 0.35$. The value $6.87$ slightly under-estimates the previously
determined fundamental frequency $\omega_f=7.04$, and results from the
positive intercept $0.35$. If we interpret the $\omega_{kf}$'s as the harmonics
of $\omega_{f}$, we should expect
$\omega_{kf} = k\omega_f$. The solid line of the upper inset of Fig.~\ref{Fig:Lomb}
shows the corresponding linear regression with the constraint of passing
through the origin.  This
gives the relation $\omega_f=6.93$, which is in good agreement
with our previous determination $\omega_f=7.04$. Another straightforward
method is to form the ratios $\omega_{kf}/k$, which should be constant
and equal to $\omega_{f}$ if the $\omega_{kf}$ are indeed the harmonics
of $\omega_{f}$. We find $\langle{\omega_{kf}/k}\rangle=7.05 \pm 0.42$,
again confirming the value of the fundamental angular log-frequency
and the presence of several harmonics.

Let us end this section by a note of caution. We have termed
``non-parametric'' the
Lomb analysis of the residues defined by (\ref{Eq:Res}). Strictly speaking,
this is not entirely correct, since the
construction of $S$ uses the fitted $t_c$, $m$, and $A$ parameters,
as criticized by \cite{F01QF}. This problem can be alleviated in several ways.
One solution is to fit the data first using a
pure power law, setting $C_1=0$ in Eq.~(\ref{Eq:LPPL}), to
obtain $A$, $m$ and $t_c$. Then, the parameters do not contain
information on the searched log-periodicity (at the linear level of description).
We have made these tests and find no differences in the results.
An alternative solution is to
estimate $A$ and $t_c$ by maximizing the linear correlation
coefficient betwen $\ln(t-t_c)$ and $y(t)-A$ and then
obtain $m$ with a simple linear regression of the two
aforementioned sequences \cite{B03QF}. In addition, one can also
perform a generalized $q$-analysis on the data $y(t)$ by scanning $t_c$
\cite{ZS02PRE}. This last method has been applied to test
log-periodicity in stock market bubbles and antibubbles
\cite{SZ02QF,ZS04IJMPC}, in the USA foreign capital inflow bubble
ending in early 2001 \cite{SZ03inflow}, and in the ongoing UK real
estate bubble \cite{ZS03PAHOUSE}. We shall use it in section \ref{s1:LPinFFR} to analyze
the log-periodic patterns of the Federal funds rate.

\section{The third-order harmonic} \label{s2:harmonic}

Previous analyses have shown that higher-order harmonics provide
significant contributions to the structure of antibubbles
\cite{SZ02QF,ZS03RG,ZS03PAGlobal}. The antibubble documented here
on the yields is no exception as we now document. Figure
\ref{Fig:Lomb} and in particular its lower inset shows that the
leading contribution to the power spectrum is provided by the
third harmonics $\omega_{3f}=3\omega_f$ rather than by the second
one $\omega_{2f}=2\omega_f$. This is reminiscent of the
log-periodicity found in two-dimensional hydrodynamic turbulence
where the strongest harmonics is also for $3\omega_f$ while
$2\omega_f$ gives almost no contribution \cite{JSH00PD}. This
suggests that, in the spirit of a Landau expansion of the type
used in \cite{SJ97land,JS99lan}, the nonlinearity is third-order
rather than quadratic.

\begin{figure}
\begin{center}
\includegraphics[width=7cm]{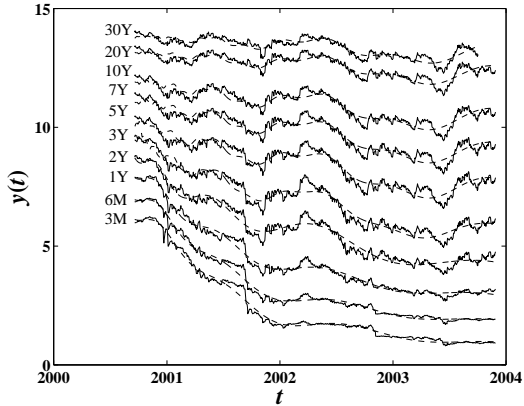}
\end{center}
\caption{Time evolution (wiggly lines) of the Yields with ten
different maturities (3M, 6M, 1Y, 2Y, 3Y, 5Y, 7Y, 10Y, 20Y, 30Y)
and their fits using formula (\ref{Eq:LPPL}) with $N=3$ and
$C_2=0$, taking into account the third-order harmonic component.
The parameters of the fits are listed in Table \ref{Tb:LPPL3}. The
yields with maturities greater than three months (3M) have been
shifted vertically upwards by 0.9\% incrementally for clarity.}
\label{Fig:YieldFit3}
\end{figure}

To test quantitatively the impact of the third harmonic, we thus
extend the previous parametric fit in section \ref{s2:Theory} by
using expression (\ref{Eq:LPPL}) with $N=3$ but fixing the
coefficient $C_2=0$ of the second harmonic. Figure
\ref{Fig:YieldFit3} shows the corresponding ten fits of the ten
yields, whose parameters are listed in Table \ref{Tb:LPPL3}.
Again, the fits to the yields with maturity larger than two years
exhibit evident LPPL signatures. The corresponding values of
$t_c$, $m$ and $\omega$ of these six fits are close to those
obtained with the simple first-order LPPL fits.  These results are
consistent with the spectral analysis reported in
Fig.~\ref{Fig:Lomb}. Since the power calculated in the Lomb
periodogram is proportional to the square of the amplitude of the
periodic component, for this parametric analysis to be consistent
with the non-parametric power spectrum, we should have
\begin{equation}
 \frac{P_N(\omega_f)}{P_N(\omega_{nf})} =
 \left(\frac{C_1}{C_n}\right)^2.
 \label{Eq:PNC}
\end{equation}
From Fig.~\ref{Fig:Lomb}, we measure
${P_N(\omega_{f})}/{P_N(\omega_{3f})}=5.8$ for the averaged Lomb
periodogram. On the other hand, $\left({C_1}/{C_3}\right)^2$ is in
the range $5-10$ for the five yields with larger maturities, using
the values reported in Table~\ref{Tb:LPPL3}. This is in reasonable
agreement with (\ref{Eq:PNC}). A similar consistency has been
reported previously for the US 2000 S\&P 500 antibubble
\cite{SZ02QF}.

\begin{table}
\begin{center}
\caption{\label{Tb:LPPL3} Parameters of fits to the ten yields shown in
Fig.~\ref{Fig:YieldFit3} using
the LPPL formula (\ref{Eq:LPPL}) with $N=3$ and $C_2=0$. }
\medskip
\begin{tabular}{ccccccccccccc}
\hline\hline
Mat&$t_c$&$m$&$\omega$&$\phi_1$&$\phi_3$&$A$&$B$&$C_1$&$C_3$&$\chi'$\\\hline
3M&1999/12/08&0.04&4.35&0.35&1.48&100.9&-77.25& 0.4760& 0.1817&0.139\\
6M&1999/12/19&-0.06&4.30&6.24&0.17&-56.32&85.93& 0.7242& 0.3329&0.132\\
1Y&2000/01/20&-0.08&4.15&4.82&2.39&-35.84&65.38& 0.5120& 0.4529&0.163\\
2Y&2000/04/15&0.26&3.51&5.44&0.48&13.19&-1.946& 0.0359& 0.0795&0.219\\
3Y&2000/10/07&0.48&7.31&3.09&2.06&6.38&-0.1542& 0.0241& 0.0044&0.244\\
5Y&2000/10/09&0.63&7.00&0.01&2.44&5.83&-0.0368& 0.0089& 0.0028&0.246\\
7Y&2000/10/24&0.66&6.80&4.84&4.20&5.81&-0.0230& 0.0070& 0.0022&0.227\\
10Y&2000/10/25&0.64&6.79&4.74&3.98&5.71&-0.0190& 0.0074& 0.0025&0.215\\
20Y&2000/10/23&0.74&6.73&4.47&2.87&5.98&-0.0052& 0.0030& 0.0012&0.183\\
30Y&2000/10/30&0.77&6.70&4.13&2.30&5.74&-0.0034& 0.0023& 0.0010&0.168\\
\hline\hline
\end{tabular}
\end{center}
\end{table}

Table \ref{Tb:LPPL3} shows some systematic effects in the variations
of the parameters $t_c$, $m$ and $\omega$ as a function of
maturity, for the six yields with maturity larger than two years.
Firstly, the
estimated critical time $t_c$ increases slightly but
systematically, suggesting an increasing (small)
lag of the antibubble inception with increasing maturity.
Secondly, the exponent $m$ increases with maturity.
Thirdly, the angular log-frequency $\omega$ decreases. We will
come back to these observations to propose an interpretation.

Is the inclusion of the third-order
log-periodic component significant? Comparing Table
\ref{Tb:LPPL1} with Table \ref{Tb:LPPL3}, we
observe that the fits including the third-order term lead to
a $\sim 10\%$ reduction in the r.m.s. of the residuals.
The fact that there is an improvement is not surprising since
there are more parameters. However, since the third-order
formula (\ref{Eq:LPPL}) with $N=3$ contains the first-order
formula  (\ref{Eq:LPPL}) with $N=1$ as a special case, we
can use the statistics of embedded hypothesis and the
 Wilks test \cite{Rao} to check
if the hypothesis $C_3=0$ can be rejected. Under the
hypothesis of i.i.d. normal residuals, the probability for
$C_3=0$ to be rejected is equal to the probability that a variable
taken from a chi-square distribution with 2 degrees of freedom
(two is the difference in the number of parameters between
the two formulas with $N=3$ and $N=1$ with $C_2=0$)
exceeds $2n\ln(\chi/\chi')$, where $n$ is the size of the time
series. For the six yields with maturities larger than 2 years, we
have $2n\ln(\chi/\chi') =$ 30, 99, 131, 146, 170, and 166,
respectively. This gives a probability that $C_3=0$ of less than
$0.01\%$. Therefore, the third-order log-periodic component appears to be
very significant.

\section{A hierarchy of antibubbles}
\label{s2:Hierarchy}

The theory underlying the LPPL formula (\ref{Eq:LPPL})
uses a so-called renormalization group formalism, which
suggests that the large scale LPPL structure studied
until now may cascade down the scales, such that LPPL
structures can be observed at many different scales
\cite{S98PR,GS02PRE,ZS03RG}. Such a
hierarchical structure with LPPL observed
as several embedded scales has been noticed over the
years by several investigators (private communications)
but was first mentioned in print at a qualitative level
in a discussion of the Deutsche Aktien Index (German stock index)
bubble before its burst in 1998 \cite{DRSW99EPJB}.
There has not been a convincing demonstration however that
the LPPL patterns can be observed unambiguously at different
time scales because small-scale structures are vulnerable to noise
\cite{SZ03QF}. The observation of LPPL at several embedded
scales become feasible when the large scale is very large, so that
the smaller scales remain large. This situation describes the
Japanese Nikkei 225 index: an antibubble was identified quantitatively
to start in 2000, that is, on top of the more-than-one-decade long
antibubble since 1990 \cite{ZS03PAGlobal}. In this case,
the small LPPL structure spans more than one year.
Actually, one year (or maybe down to six
months) might be the lower cutoff for the detection of
statistically significant LPPL structures in financial bubbles or antibubbles,
based on daily data. Johansen proposed that LPPL patterns
can only be ascertained about a time scale of
one or two years \cite{J03QF}. Here we provide
yet another example of such LPPL-within-LPPL structure for the
yield data.

\begin{figure}[h]
\begin{center}
\includegraphics[width=7cm]{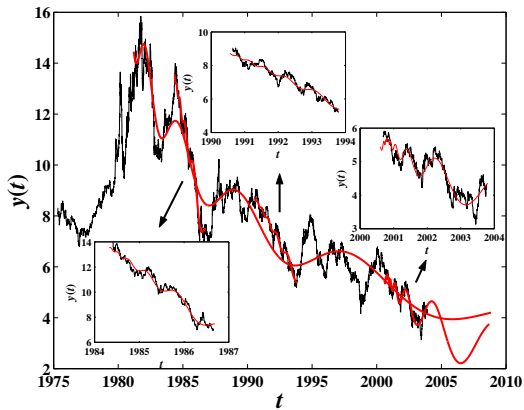}
\end{center}
\caption{Illustration of the concept of a hierarchy of
LPPL structures in antibubbles. The figure shows a large-scale
antibubble on the U.S. 10-year treasury bond yield from 1979 to 2003.
In addition , three small-scale antibubbles are detected as shown in the
insets which magnify the three small-scale
antibubbles. The smooth oscillatory
lines show the fits to a large-scale antibubble and three
small-scale antibubbles.} \label{Fig:Hierarchy}
\end{figure}

Figure \ref{Fig:Hierarchy} shows 29 years of the evolution of the
U.S. 10-year treasury bond yield from 1975 to 2003. It is
interesting to realize that the evolution at such large time
scales can be represented also by an antibubble that started in
1979. The inception of this antibubble of the yield
is associated with the peak in 1979-1980 of the inflation rate.
Figure \ref{Fig:Hierarchy} also shows three small-scale
antibubbles embedded within the large-scale.
We have fitted these four antibubbles with the
first-order LPPL formula (\ref{Eq:LPPL}). The data set used in these
fits goes from the local high to the local
low that breaks the structure when available \cite{JSL99JR}.
The smooth oscillatory lines in
Fig.~\ref{Fig:Hierarchy} show the fits to a large-scale antibubble
and three small-scale antibubbles.

The parameters of the four fits are the following. The data set of
the large-scale antibubble used in the LPPL fit is from 1981/09/30 to
2003/10/03 and we find $t_c={\rm{1979/01/04}}$, $m=-0.17$,
$\omega=10.36$, $\phi=0.52$, $A=-15.27$, $B=98.5917$, $C=-3.4737$,
with a r.m.s. of the fit residuals equal to $\chi=0.711$. The data set
for the first small-scale antibubble used in the LPPL fit goes from 1984/05/30 to
1986/08/29 and we find $t_c={\rm{1984/02/23}}$, $m=1.04$,
$\omega=12.41$, $\phi=4.20$, $A=14.01$, $B=-0.0057$, $C= 0.0006$,
with a r.m.s. of the fit residuals equal to $\chi=0.323$. The data set
of the second small-scale antibubble goes from 1990/08/24 to
1993/10/15 and we find $t_c={\rm{1989/10/28}}$, $m=0.99$,
$\omega=11.19$, $\phi=2.10$, $A=9.49$, $B=-0.0027$, $C= 0.0002$,
with the r.m.s. of the fit residuals equal to $\chi=0.233$. The data set
of the third small-scale antibubble goes from 2000/09/09 to
2003/10/10 and we find $t_c={\rm{2000/10/23}}$, $m=0.62$,
$\omega=7.03$, $\phi=3.18$, $A=5.72$, $B=-0.0213$, $C=-0.0089$,
with a r.m.s. of the fit residuals equal to $\chi=0.237$. This last antibubble
has already been documented in Table \ref{Tb:LPPL1}.

\begin{figure}
\begin{center}
\includegraphics[width=8cm]{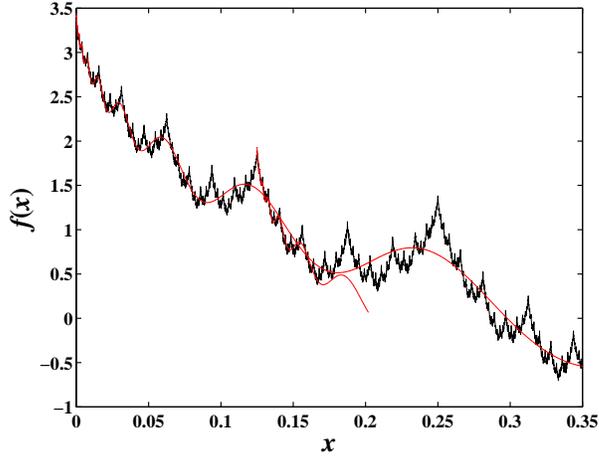}
\end{center}
\caption{Weierstrass-Mandelbrot function \cite{Singh,Berry,GS02PRE}
as a model  of antibubbles within antibubbles within ...
The Weierstrass-Mandelbrot function is
defined as $f(x) = \sum_{n=0}^{N} b^n \cos(a^n\pi x)$ with
$N\to \infty$. The parameters used for
this figure are $a = 2$ and $b = 1/2^0.5$. The two continuous lines
show the fit of the Weierstrass-Mandelbrot function with
equation (\ref{Eq:LPPL}) with $N=1$ both globally (large-scale
antibubble) and locally (immediately smaller-scale antibubble).
There are many more embedded smaller scales (actually an infinity
when infinite resolution is available), corresponding to a self-similar
fractal function. \protect\label{WM1}}
\end{figure}

\begin{figure}
\begin{center}
\includegraphics[width=8cm]{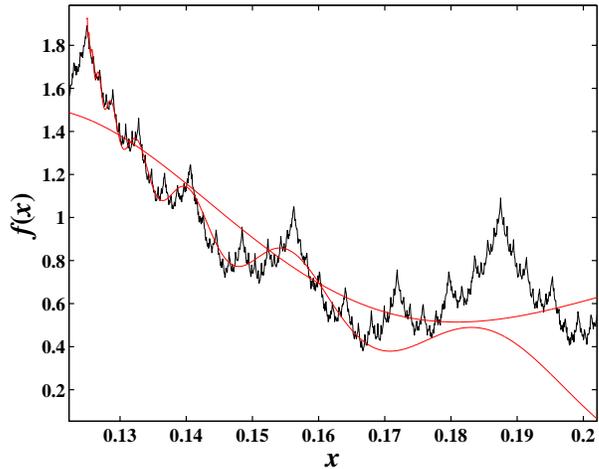}
\end{center}
\caption{Magnification of Fig.~\ref{WM1}. \protect\label{WM2}}
\end{figure}

The 1979-2003 large-scale antibubble and the recent 2000-2003
small-scale antibubble provide predictions for the future
evolution of the 10-year yield, obtained by extrapolating the LPPL
fits to the coming years. The extrapolation of the large-scale
antibubble suggests that the 10-year yield will decrease for about
two years and then rebound, while that of the small-scale
antibubble suggests an increase following by a reversal during the
first quarter of 2004 which could last two years before reverting
to growing again. At face value, it seems that these two
predictions are contradictory: they can not be both right. This is
actually not the case: the essence of the renormalization group
formulation of antibubbles is that small structures can be carried
by larger structures, in the way exemplified in Figs.~\ref{WM1}
and \ref{WM2}. The small-scale structure is carried by the
large-scale structure and provides details of the overall pattern
delineated by the large scale structure. Thus, the two scenarios
shown in Fig.~\ref{Fig:Hierarchy} can be reconciled by not
opposing them but by combining them and viewing them as two
predictions of the same overall multi-scale process observed at
two different time resolutions. Combining the prediction of these
two time scales suggest that the 10-year yield will decrease at
large scale with local rallies and falls until its recovery two
years later.

\section{Log-periodic patterns in Federal funds rate}
\label{s1:LPinFFR}

It is well-known that the yields are driven by the Federal funds
rate, which fixes the target overnight interest rate. The driving
of the yields with longer maturities by the Federal funds rate is
transmitted by the ``rigidity'' of the yield curve (or similarly
of the forward rate curve) controlled by arbitraging
\cite{string1,string2}. A simple picture is to imagine an elastic
string whose handle (the Federal funds rate) is held by the
Federal Reserve Chairman who moves it up or down, while the rest
of the string slowly moves with lags and delays due to the
interplay of inertial and elastic interactions.

In view of our finding of significant LPPL structures
in yields with maturities larger than 2 years, the question
naturally arises as to their origin. Are the LPPL structures
intrinsic or endogenous to these yields with medium and large
maturities? Or are they reflecting and amplifying a log-periodicity
already present in the driving Federal
funds rate? The difficulty of this question lies in the fact
that the Federal funds rate is a staircase of plateaus with jumps
and the number of jumps is not large. In a first bold attempt,
we have fitted the Federal funds rate with the LPPL
formula (\ref{Eq:LPPL}) following the same procedure used for
fitting the other yields. We are not able to find a reliable
signal because the power law part of the formula is absent.
But, equation (\ref{Eq:LPPL}) contains two ingredients: (i)
a power law $(t-t_c)^m$ and a log-periodic oscillation
$\cos [\omega \ln(t-t_c)]$.
Is it still possible for the log-periodic pattern to be
present while the power law is absent? To address this question,
we note that this
amounts to asking if the times $t_n$ at which the funds rate
is changed by the Federal Reserve could exhibit a geometric time
series in the variable $t_n-t_c$ with a suitable value of $t_c$.

To our knowledge, the most robust method to attack this question
is to use the generalized
$(H,q)$-analysis on the Federal funds rate $y(t)$. The
$(H,q)$-analysis \cite{ZS02PRE,ZS04IJMPC} is a generalization of
the $q$-analysis \cite{E97PLA,EE97PRL}, which is a natural tool
for the description of discretely scale invariant fractals. The
$(H,q)$-derivative of the function $y(\tau)$ is defined as
\begin{equation}
D_q^H y(\tau) \stackrel{\triangle}{=} \frac
{y(\tau)-y(q\tau)}{[(1-q)\tau]^H}~, \label{Eq:HqD}
\end{equation}
where $\tau = t-t_c$ is the time to the critical value $t_c$. The special
case $H=1$ recovers the normal $q$-derivative, which itself
reduces to the normal derivative in the limit $q \to 1^-$. There
is no loss of generality by constraining $q$ in the open interval
$(0,1)$ \cite{ZS02PRE}. The parameter $q$ tests for the
log-periodic structure, while $H$ acts as a high-pass filter to
remove the global trend in the time series ensuring that the
resultant series is almost stationary. The introduction of $H$
allows us to analyze non-fractal signals which nevertheless
contain log-periodic components.

In order to perform the $(H,q)$-analysis of the Federal funds rate
$y(t)$, we need to choose a value for $t_c$. In this goal, we use
the following guidelines. First, Table \ref{Tb:LPPL3} shows that
$t_c$ for the yields decreases when the maturity decreases.
Extrapolating, this suggests that the $t_c$ for the Federal funds
rates is prior (but probably not much) to those estimated for the
yields with larger maturities. This should be the case if the main
driver of the yield rates was fixed by commercial banks who react
following the central bank. It turns out that the $(H,q)$-analysis
is very robust with respect to misspecification of $t_c$: the
absence of an accurate estimate of $t_c$ does not impact much on
the extraction of log-periodic components if (1) $q$ is close to
$1$ so that the analysis is performed at a finer resolution and
(2) $H$ is negative so that the part of $(H,q)$-derivative far
from the critical time $t_c$ pays a more importance role in the
analysis. In practice, using positive $H$'s give a majority of the
angular log-frequencies with very low values: this can be
interpreted to be caused by the remaining global trend in the
signal. Also, if we use small $q$'s, we find small and unstable
angular log-frequencies.

\begin{figure}[h]
\begin{center}
\includegraphics[width=7cm]{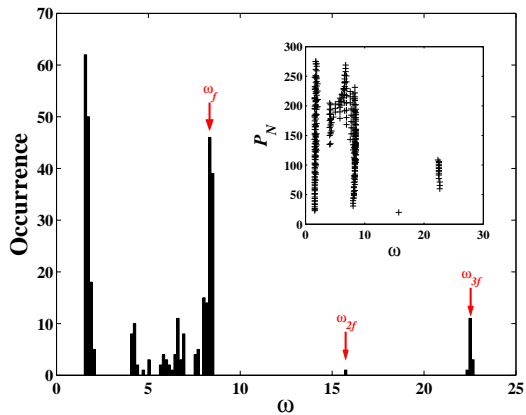}
\end{center}
\caption{Histogram of the angular log-frequency $\omega$ of the
most significant Lomb peaks in the generalized $(H,q)$-analysis of the
Federal funds rate. The
inset shows the bivariate distribution of the pairs $(\omega,P_N)$
of the angular log-frequency at the largest peak of the Lomb
periodograms, together with its amplitude $P_N$.}
\label{Fig:HqA}
\end{figure}

In order to implement concretely the $(H,q)$-analysis, we use here
$t_c={\rm{2000/10/20}}$, which is compatible with the above
considerations. We scan a $21 \times 16$ rectangular grid in the
$(H,q)$ parameter plane, with $H = -1:0.1:1$ and $q =
0.6:0.02:0.9$. For each pair of $(H,q)$, we calculate the
$(H,q)$-derivative, on which we perform a Lomb periodogram
analysis. We have found that most of the Lomb periodograms have a
similar shape as shown in Fig.~\ref{Fig:Lomb}. The highest Lomb
peak of the resultant periodogram allows us to identify an angular
log-frequency $\omega$ with its height $P_N$, which are both
functions of $H$ and $q$. Figure~\ref{Fig:HqA} shows the histogram
of occurrences of $\omega$. Its inset gives the bivariate
distribution of pairs $(\omega,P_N)$. There are clear clusters.
The cluster peaked at low $\omega$ close to 1.7 is produced by
relatively large (usually positive) $H$'s and relatively small
$q$'s. The part of the histogram for $\omega<8$ corresponds to
$q<0.76$. For $q>0.76$ and negative $H$'s, we see a well-defined
cluster at $\omega$ close to $8.3\pm 0.2$ (indicated by an arrow
with $\omega_f$). We attribute this value to the fundamental
$\omega_f$. We can also discern two harmonics at $\omega_{2f}
\approx 15.7$ and $\omega_{3f} \approx 22.5$, indicated by arrows.

We thus conclude that there is no power law but a significant
log-periodic structure in the Federal funds rate since October 2000,
with an estimated
angular log-frequency $\omega_f = 8.3\pm 0.2$.

Theoretically, the $(H,q)$-analysis can be employed to estimate
$t_c$ endogenously by searching for the highest Lomb peak obtained
by varying $t_c$ in some range. However, the predictive power of
this method has proven rather weak in other applications
\cite{ZS02PRE,ZS04IJMPC} and we do not pursue it here. Changing
$t_c$ by a few months both ways does not change significantly our
conclusions.

\section{Causality between stock markets and Federal funds
rate} \label{s1:Causality}

The similarity between the LPPL structures observed in the stock market
since August 2000 and in the yields since October 2000 begs
the question of the origin of these common properties. Let us consider the
following scenarios.
\begin{enumerate}
\item Is the FRB (reacting to macroeconomic indicators together
with its monetary policy) the source of log-periodicity in the
timing of its interventions, which then spills over and is
amplified in the yields with larger maturities, themselves
influencing the stock market? In this scenario, the FRB is the
source of log-periodicity in its response to other stimuli, which
is then transferred to bonds and then to stocks.

\item Is the stock market first developing LPPL which then influence
the yields of large maturities, both of them influencing the FRB
timing of its intervention?

\item Or, keeping the hypothesis of the primary influence of the LPPL of
the stock market, is the FRB the secondary step in reacting to the
stock market which then
spills over and influences the yields with larger maturities?

\item Is the LPPL observed both on the stock market and the yields the
result of an overarching common origin to which they respond almost
simultaneously?
\end{enumerate}
In short, we address a problem of causality between two or more time series,
which has a rich literature in economics (see for instance
\cite{Granger,Geweke} and references therein). Determining causality
is notoriously difficult and often ill-defined and one has in general
to assume drastic simplifications, such as stationarity of the time series,
linear least-squares projections, and mean-square errors.
While these assumptions are convenient to
make when conducting empirical tests, they are not realistic.
Here, we draw from our analysis to propose that the most probable scenario
is number 2: Stock Market $\to$ Fed Reserve (Federal
funds rate) $\to$ short-term yields $\to$ long-term yields.
This conclusion is based on the following.

First, we have found that the critical inception times $t_c$
of the yields as well as of the Federal funds rate are
about two months after that of the
2000 US stock market antibubble. In addition, $t_c$ tends to
increase with the yield maturity.

Second, we notice that the log-periodic angular frequency
decreases from the value $\omega = 10.3 \pm 0.2$ for the S\&P 500
Index to $\omega = 8.3 \pm 0.2$ for the
Federal funds rate and to values $7< \omega<8$ for yields
when increasing their maturity (see Table \ref{Tb:LPPL3}).
This trend in $\omega$ is consistent with the view that
the Federal funds rate cuts was driven by the US stock market
oscillations and thus developed a lower log-periodic frequency,
which in turn drove the yields into still lower log-periodic frequencies.

\begin{figure}[h]
\begin{center}
\includegraphics[width=7cm]{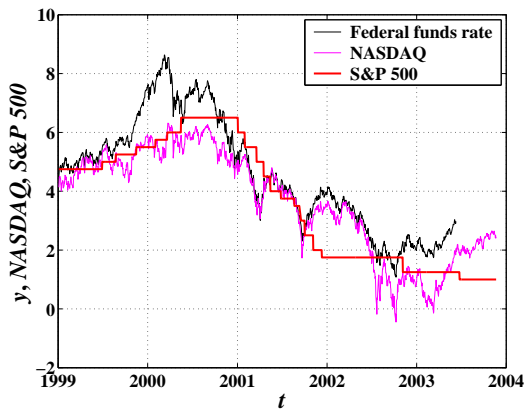}
\end{center}
\caption{(Color online) Comparison of the Federal funds rate, the
S\&P 500 Index $x(t)$, and the NASDAQ Composite $z(t)$, from 1999
to mid-2003. To allow a illustrative visual comparison, the
indices have been translated and scaled as follows: $x \to 5x-34$
and $z \to 10z-67$.} \label{Fig:FFRSP}
\end{figure}

Third, Fig.~\ref{Fig:FFRSP} shows in the same plot the Federal
funds rate $y(t)$, the logarithm of the S\&P 500 Index $x(t)$, and
the logarithm of the NASDAQ Composite $z(t)$. One observes that
the Federal funds rate $y(t)$ was increasing in the bull market
and has been decreasing in bear market. After the burst of the new
economy bubble, the Federal Reserve cut the interest rate from
$6^{\frac{1}{2}}$ to 6 on 2001/01/03, at a time lagging more than
four months behind the onset of the stock market antibubble and
more than eight months after the burst of the bubble. This
suggests that this Federal Reserve's interest rate moves in the
recent period have been caused at least partially by the behavior
of the stock markets.

\begin{figure}[h]
\begin{center}
\includegraphics[width=7cm]{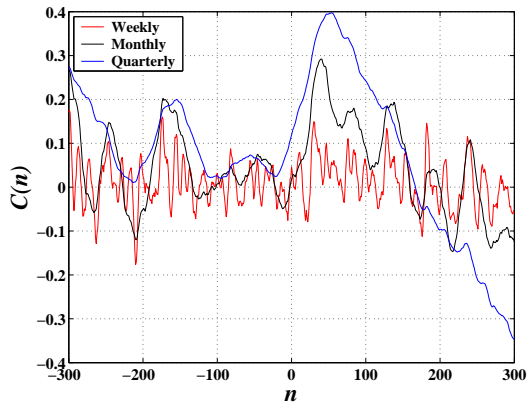}
\end{center}
\caption{(Color online) Cross-correlation coefficient $C(n)$
between the increments of the logarithm of the S\&P 500 Index and
the increments of the Federal funds rate as a function of time lag
$n$ in days. The three curves corresponds to three different time
steps used to calculate the increments: weekly, monthly and
quarterly. A positive lag $n$ corresponds to having the Federal
funds rate posterior to the stock market.} \label{FigCorrOct1999}
\end{figure}

Fourth, Fig.~\ref{FigCorrOct1999} presents the cross-correlation
between the Federal funds rate and the S\&P 500 Index in the time
period from October 1999 to mid-2003. The results are essentially
the same when varying the start date from October 1999 to October
2000. Specifically, we construct the increments $\Delta y(t)$ of
the Federal funds rate and of the logarithm of the S\&P 500 Index
(which defines the returns $\Delta \ln x(t)$) at the weekly,
monthly and quarterly scales. Taking increments ensure to work
with time series which are approximately stationary. We then
calculate the cross-correlation coefficient with the Federal funds
rate translated by $n$ time steps (trading days) according to the
formula
\begin{equation}
C(n) = {\rm{Corr}}\left(\Delta \ln x(t), \Delta y(t+n)\right)~,
\label{Eq:Cn}
\end{equation}
where ${\rm{Corr}}(x,y)$ is the statistical correlation between
$x$ and $y$. Figure~\ref{FigCorrOct1999} shows the
cross-correlation coefficient function $C(n)$ as a function of
$n\in[-300,300]$ for the three time scales. While the weekly scale
is too noisy to conclude, the cross-correlation coefficient
exhibits a clear maximum for the two other time scales for a
positive lag in the range 30-50 trading days: such a positive lag
means that the S\&P 500 Index in the past has a predictive power
on the Federal funds rate in the future. Since correlations detect
only linear predictability, this means that there is a linear
(Granger) predictability of the Federal funds rate by the stock
market.

\begin{figure}[h]
\begin{center}
\includegraphics[width=7cm]{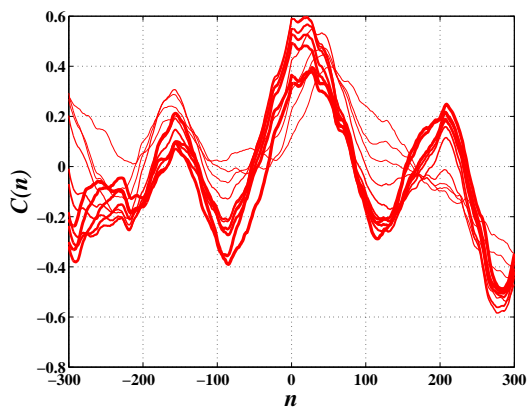}
\end{center}
\caption{Each curve shows the cross-correlation coefficient $C(n)$
between the increments of the logarithm of the S\&P 500 Index and
the increments of the lagged bond yield at a given maturity as a
function of time lag $n$. There are 11 curves with increasing line
width corresponding to the maturities 0M (Federal funds rate), 3M,
6M, 1Y, 2Y, 3Y, 5Y, 7Y, 10Y, 20Y and 30Y. The increments have been
calculated using a time scale of one month. A positive lag $n$
corresponds to having a bond yield posterior to the stock market.}
\label{FigXCorrOct1999}
\end{figure}

Figure~\ref{FigXCorrOct1999} is similar to
Fig.~\ref{FigCorrOct1999} with the following modifications: it
uses a single time scale of one quarter to calculate the
increments and shows the cross-correlation coefficient between the
increments of the logarithm of the S\&P 500 Index and the
increments of the lagged yields for all maturities 0M (Federal
funds rate), 3M, 6M, 1Y, 2Y, 3Y, 5Y, 7Y, 10Y, 20Y and 30Y.

Fig. \ref{FigXCorrOct1999} exhibits the robust feature that the
cross-correlation coefficients of the pairs (stock market; bond
yield) for the smallest maturities 0M (Federal funds rate), 3M,
6M, 1Y peak for a lag close to 30-50 days. In contrast, the yields
with longer maturities have their cross-correlation with the S\&P
500 Index exhibit two peaks, one for a lag around 30 days and one
for a zero lag. This gives us an important information in addition
to that obtained from Fig.~\ref{FigCorrOct1999}: not only do we
see the impact of the S\&P 500 Index in the past on all future
bond yields, we also find that the S\&P 500 Index is impacting or
causing the variation of the bond yield with large maturities
instantaneously. This is consistent with the scenario in which the
S\&P 500 Index ``causes'' the FRB moves which itself drives the
yields with the shorter maturities and at the same time the yields
with the longer maturities are directly and instantaneously
influenced by the S\&P 500 Index (and the stock market more
generally). In other words, we are uncovering the existence of two
driving forces on the yields: the stock market and the monetary
policy of the FRB. The stock market prevails strongly for
long-term yields and in addition influences significantly causally
the FRB actions which then dominates in its influence on the
short-term yields.

The existence of a positive lag found here means that the S\&P 500
Index in the past has a linear Granger predictive power on bond
yields in the future for all maturities In other words, we can
assert that there is a Granger causality of the stock market on
the bond yields, at least in the recent past, from 2000 till
mid-2003.

\section{Conclusion}
\label{s1:concude}

Using the descriptive method of log-periodic power laws (LPPL)
based on a theory of behavioral herding, we have shown the
existence of an antibubble in the yields of maturities larger than 1 year
since October 2000. The concept of ``antibubble'' describes the
existence of a specific LPPL pattern that is thought to reflect
collective herding effects. We have presented a series of tests on
the significance of log-periodicity: based on the existence of
strong harmonics, on the fact that the dominant harmonics improve
very significantly the power of explanation of the yield data, on
the consistency between our parametric and non-parametric analyses,
we can assert the existence of the LPPL antibubble on the yields since
October 2000. We have then performed a simple cross-correlation
calculation which has confirmed the following scenario, already
suggested from the dependence of the parameters of the LPPL formula
as a function of yield maturities:
\begin{itemize}
\item Herding and competition with value investing has led to a
collective LPPL antibubble unfolding in the U.S. stock markets
since August 2000 (and which may have already ended
\cite{Zhousor_end}). \item The Federal Reserve Board (FRB) reacted
to and lagged (by about 30 trading days) behind
 the log-periodicity of the stock markets by a series
of steps lowering the Federal funds rate according to a pattern
mimicking the log-periodicity of the stock market.
\item  The yields with maturities less than 1 or 2 years
then followed by amplifying the FRB steps by the interplay of
risk and herding.
\item The yields with larger maturity were also influenced
directly and instantaneously by the stock market.
\end{itemize}
Our interpretation is that the FRB is ``causally slaved'' to the
stock market, because the later is now more and more
considered as a proxy for the present and
future health of economy. The impact of the stock market both
economically and psychologically is obvious from the large
fraction of the US population actively invested and by the sheer
size of the stock market capitalization that has grown
significantly larger than the GDP.

The present work reinforces the picture in which anticipations and collective beliefs
play growing roles in controlling and shaping the future evolution of
not only the stock market but also the economy itself.

\textbf{Acknowledgments}

We are grateful to P.M. Thomas for attracting our attention
to this problem. This work was
supported by the James S. Mc Donnell Foundation 21st century
scientist award/studying complex system.


\end{document}